\documentclass[10pt,a4paper,journal]{IEEEtran}

\usepackage[a4paper,
            left=0.6in,right=.5in,top=0.5in,bottom=0.75in]{geometry}
\usepackage{caption}

\usepackage{graphics}
\usepackage{epsfig}
\usepackage{graphicx}
\usepackage{caption}
\usepackage{color}
\usepackage{amssymb}
\usepackage{amsthm}
\usepackage{algorithm}
\usepackage{algorithmic}
\usepackage{amsmath}

\usepackage{cite}
\usepackage{url}
\usepackage{color}
\usepackage[caption=false]{subfig}
\usepackage{amssymb}
\usepackage{amsmath}
\usepackage{commath}
\usepackage{graphicx,bm}
\usepackage{verbatim}
\newcommand{\subparagraph}{}
\usepackage{multirow}

\newcounter{tempEquationCounter}
\newcounter{thisEquationNumber}
\newenvironment{floatEq}
{\setcounter{thisEquationNumber}{\value{equation}}\addtocounter{equation}{1}
\begin{figure*}[!t]
\normalsize\setcounter{tempEquationCounter}{\value{equation}}
\setcounter{equation}{\value{thisEquationNumber}}
}
{\setcounter{equation}{\value{tempEquationCounter}}
\hrulefill\vspace*{4pt}
\end{figure*}

}

\usepackage{amssymb}
\usepackage{amsfonts}
\usepackage{lineno}
\usepackage{comment}

\newtheorem{definition}{\bf Definition}

\newcommand{\sue}{c} 
\newcommand{\us}{l} 
\newcommand{\pc}{p}
\newcommand{\op}{\beta} 
\newcommand{\EP}{\zeta} 

\usepackage{paralist} 

\usepackage{verbatim} 

\begin{document}
\setlength{\parskip}{0cm}
\IEEEpubid{\makebox{ISBN 978-3-8007-4221-9 (c) © VDE VERLAG GMBH, Berlin, Offenbach, Germany \hfill }
\hspace{\columnsep}\makebox{\hfill }}

\title{Joint In-Band Backhauling and Interference Mitigation in 5G Heterogeneous Networks}

\author{Trung Kien Vu,~\IEEEmembership{Student Member,~IEEE,}
        Mehdi Bennis,~\IEEEmembership{Senior Member,~IEEE,}\\
        Sumudu~Samarakoon,~\IEEEmembership{Student Member,~IEEE,}
        M\' erouane Debbah,~\IEEEmembership{Fellow,~IEEE,}
        and Matti Latva-aho,~\IEEEmembership{Senior Member,~IEEE,}

\thanks{Manuscript received March 15, 2016; revised and accepted April 19, 2016. Date of publication May 20, 2016; This work was supported in part by the Finnish Funding Agency for Technology and Innovation (Tekes),  Nokia,  Huawei,  Anite, in part by the Academy of Finland funding through the grant 284704, and in part by the ERC Starting Grant 305123 MORE (Advanced Mathematical Tools for Complex Network Engineering).}

\thanks{T. K. Vu, M. Bennis, S.~Samarakoon, and M. Latva-aho are with the Centre for Wireless Communications, University of Oulu, Oulu 90014, Finland (e-mail: trungkien.vu@oulu.fi; mehdi.bennis@oulu.fi; sumudu.samarakoon@oulu.fi; matti.latva-aho@oulu.fi).}

\thanks{M. Debbah is with the Large Networks and System Group (LANEAS), CentraleSup\'elec, Universit\'e Paris-Saclay, Gif-sur-Yvette, France and is with the Mathematical and Algorithmic Sciences Laboratory, Huawei France R\&D, Paris, France  (e-mail: merouane.debbah@huawei.com).}

\thanks{Citation information: In proceeding of 2016 IEEE European Wireless} }\vspace{-6.0cm}

\maketitle

\begin{abstract}
In this paper, we study the problem of joint in-band backhauling and interference mitigation in $5$G heterogeneous networks (HetNets) in which a massive multiple-input multiple-output (MIMO) macro cell base station equipped with a large number of antennas, overlaid with self-backhauled small cells is assumed. This problem  is cast as a network utility maximization subject to wireless backhaul constraints. Due to the non-tractability of the problem, we first resort to random matrix theory to get a closed-form expression of the achievable rate and transmit power in the asymptotic regime, i.e., as the number of antennas and users grows large. Subsequently, leveraging the framework of stochastic optimization, the problem is decoupled into dynamic scheduling of macro cell users and backhaul provisioning of small cells as a function of interference and backhaul links. Via simulations, we evaluate the performance gains of our proposed framework under different network architectures and low/high frequency bands. Our proposed HetNet method achieves the achievable average UE throughput of $1.7$ Gbps as well as ensures $1$ Gbps cell-edge UE throughput when serving $200$ UEs per $\text{km}^{2}$ at $28$ GHz with $1$ GHz bandwidth. In ultra-dense network, the UE throughput at $28$ GHz achieves $62\times$ gain as compared to $2.4$ GHz. Moreover, we show that there exists an $[\mathcal{O}(1/\nu), \mathcal{O}(\nu)]$ utility-queue backlog tradeoff with any arbitrary non-negative $\nu$.

\end{abstract}

\IEEEpeerreviewmaketitle
\section{Introduction}
\label{Intro}
The demand for massive data traffic has grown due to the exponential increase in the number of mobile broadband subscribers such as smartphones and tablets. To handle these relentless demands, in the next generation wireless networks, a number of candidate solutions include: $1$)  higher frequency spectrum (centimeter and millimeter-wave (mmWave)); $2$) advanced spectral-efficiency techniques (massive MIMO); and $3$) ultra-dense small cell deployments~\cite{Nokia2011}. Massive MIMO plays an important role in wireless networks due to an improvement in energy and spectral efficiency~\cite{marzt2010non}. The basic concept of massive MIMO assumes a macro base station (MBS) equipped with a few hundreds antennas simultaneously serving tens of macro user equipments (MUEs). On the other hand, ultra dense SC deployment provides an effective solution to increase network capacity. In parallel to that, recent advances in full-duplex (FD) enables doubling spectral efficiency and lowering latency, which is another promising technique~\cite{li2015small}.

The combination of massive MIMO and ultra-dense SCs holds the promise of ensuring high capacity improvement and constitutes the main motivation of this work. Recently, SC wireless backhaul in massive MIMO was studied in~\cite{li2015small, S2014if} showing the efficiency of combining massive MIMO and wireless backhaul-based small cell networks focusing on minimizing power consumption. However, the key challenge of how to dynamically optimize
the overall network performance taking into account the backhaul dynamics, and scheduling has not been addressed. The main contribution of this work is to study the problem of joint scheduling, interference mitigation, and in-band wireless backhauling. We design precoders to alleviate both co-tier and cross tier interference and dynamically provide wireless backhauling to SCs. An operation mode is proposed to dynamically control the interference among adjacent SCs. By invoking results from random matrix theory (RMT), we derive a closed-form expression of the signal-to-interference-plus-noise-ratio ($\mathrm{SINR}$) when the numbers of MBS antennas and users grow very large. A network utility optimization problem is formulated to maximize the total network throughput subject to dynamically varying wireless backhaul. Leveraging the framework of stochastic optimization, the problem is decoupled into several subproblems. The mixed-integer non-convex subproblem is solved by applying the framework of successive convex approximation (SCA). A performance evaluation is carried out for different network deployments at low/high frequency bands.
\IEEEpubidadjcol

The rest of this paper is organized as follows. Section~\ref{lb-SM-P} describes the system model and Section~\ref{Op-Form} formulates the problem of scheduling and interference mitigation. Section~\ref{LOF} adopts the Lyapunov optimization framework and the SCA method used to solve the problem. In Section~\ref{Evaluation}, the simulation results are presented and Section~\ref{Conclusion} concludes the paper.
\section{System Model}
\label{lb-SM-P}
\subsection{System Model} 
\label{lb-SystemModel}
\begin{figure}[!h]
\begin{minipage}[t]{1\linewidth}
    \centering
    \includegraphics[scale=0.5]{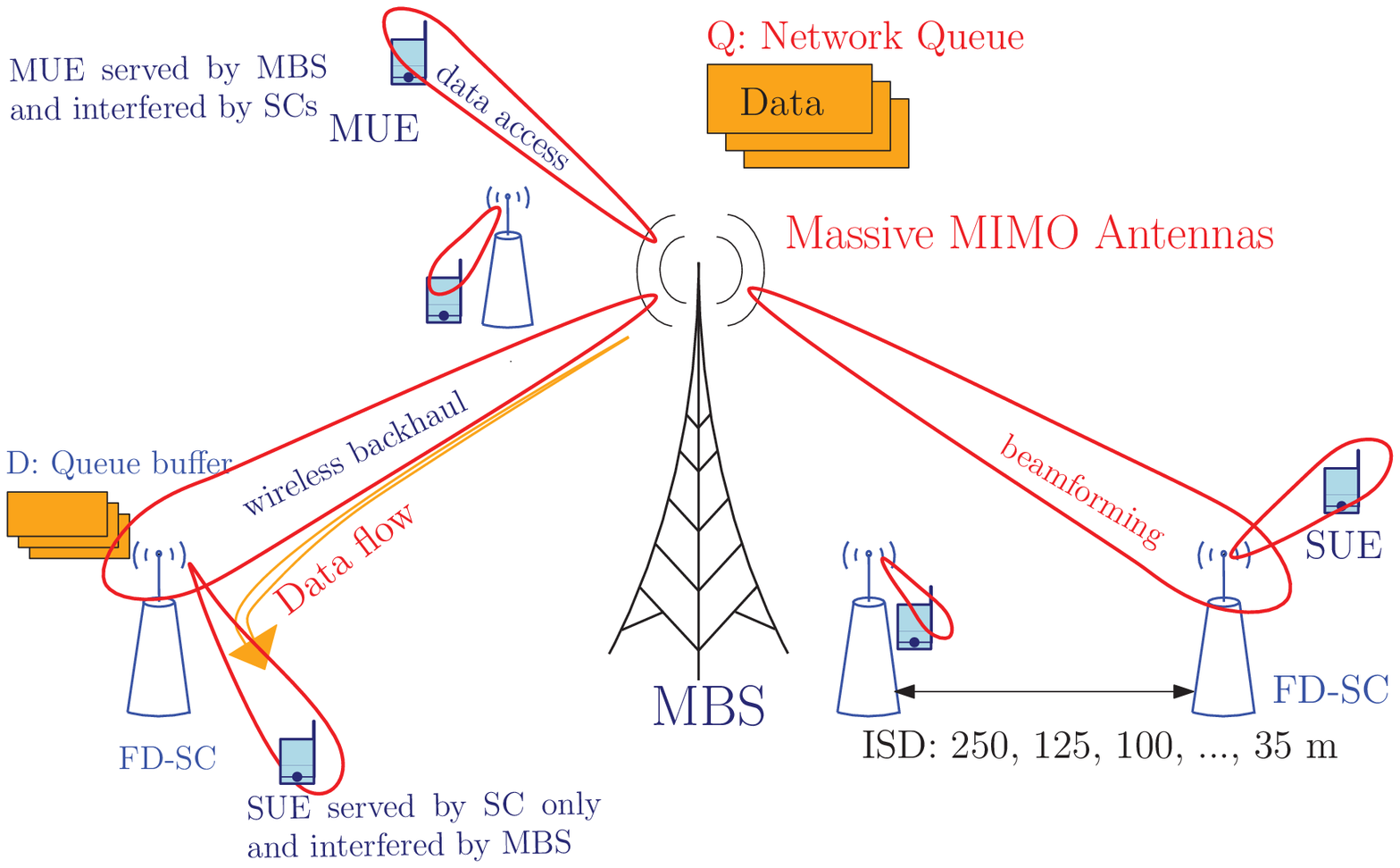}
    \caption{Integrated Access and Backhaul Architecture for 5G network.}\label{Topology}
\end{minipage}
\end{figure}

We consider the downlink (DL) transmission of a HetNet scenario as shown in Fig.~\ref{Topology} in which a MBS $b_{0}$ is underlaid
with a set of uniformly deployed $\emph{S}$ SCs, $\mathcal{S} = \{b_{s}| s \in \{1, \ldots, \emph{S}\}\}$. Let $\mathcal{B} = \{b_{0}\} \cup \mathcal{S}$ denote the set of all base stations (BSs), where $|\mathcal{B}| = 1 +  \emph{S}$. The MBS is equipped with $\emph{N}$ number of antennas and serves a set of single-antenna $\emph{M}$ MUEs $\mathcal{M} = \{1, \ldots, \emph{M}\}$. Let $\mathcal{K} = \mathcal{M} \cup \mathcal{S}$ denote the set of $b_0$'s associated users, where $|\mathcal{K}| =  \emph{K} = \emph{M} +  \emph{S}$. SCs are assumed to be FD capable with perfect self-interference cancelation (SIC) capabilities\footnote{The case of imperfect SIC is left for future work.}. Each SC is equipped with two antennas: the receiving antenna is used for the
wireless backhaul and the transmitting antenna to serve its single-antenna small cell user equipment (SUE)\footnote{In general, one SC base station can serve more
than one SUE.Without loss of generality, one SUE with higher traffic demand can be considered as multiple
SUEs with lower traffic demand. More SUEs per SC is left for future work.}. Let $\mathcal{C}=\{ {\sue_{1}}, {\sue_{2}}, \ldots, {\sue_{S}} \}$ denote the set of SUEs, where $|\mathcal{C}| = \emph{S}$. We assume closed-access policy where SCs serve their own users. Co-channel time-division duplexing (TDD) protocol is considered in which the MBS and SCs share the entire bandwidth, and do the DL transmission at the same time. In this work, we consider a large number of antennas at the MBS and a dense deployment of MUEs and SCs, such that $\emph{M}, \emph{N},\emph{S} \gg 1$.
\subsection{Channel Model}
\label{lb-ChannelModel}
We denote $\mathbf{h}_{m}^{(b_{0})} = \big[ h_{m}^{(b_{0},1)}, h_{m}^{(b_{0},2)}, \cdots, h_{m}^{(b_{0},\emph{N})} \big]^{T} \in \mathbb{C}^{\emph{N} \times 1}$ the propagation channels between the $m{\text{th}}$ MUE and the antennas of the MBS $b_{0}$ in which $h_{m}^{(b_{0},n)}$ is the channel between the $m{\text{th}}$ MUE and $n{\text{th}}$ MBS antenna. Let $\mathbf{H}^{(b_{0}), \text{M}} = \big[\mathbf{h}_{1}^{(b_{0})}, \mathbf{h}_{2}^{(b_{0})}, \cdots,
\mathbf{h}_{\emph{M}}^{(b_{0})} \big] \in \mathbb{C}^{\emph{N} \times \emph{M}}$ denote the channel matrix
between all MUEs and the MBS antennas. Moreover, we assume imperfect channel state information (CSI) for MUEs due to mobility and we denote
$\mathbf{\hat{H}}^{(b_{0}), \text{M}} = \big[ \mathbf{\hat{h}}_{1}^{(b_{0})}, \mathbf{\hat{h}}_{2}^{(b_{0})}, \cdots,
\mathbf{\hat{h}}_{\emph{M}}^{(b_{0})}\big] \in \mathbb{C}^{\emph{N} \times \emph{M}}$ as the estimate of
$\mathbf{H}^{(b_{0}), \text{M}}$ in which the imperfect CSI can be modeled as~\cite{rusek2013s}:
\begin{equation} \label{eq:h-MUE}
 \mathbf{\hat{h}}_{m}^{(b_{0})} = \sqrt{\emph{N} \mathbf{\Theta}_{m}^{(b_{0})} } \mathbf{\hat{w}}_{m}^{(b_{0})},
\end{equation}
where $\mathbf{\hat{w}}_{m}^{(b_{0})} = \sqrt{1 - {\tau_m}^2} \mathbf{w}_{m}^{(b_{0})} + \tau _m \mathbf{z}_{m}^{(b_{0})}$ is the estimate of the small-scale fading channel matrix and $\mathbf{\Theta}_{m}^{(b_{0})}$ is the spatial channel correlation matrix that accounts for path loss and shadow fading. Here, $\mathbf{w}_{m}^{(b_{0})}$ and $\mathbf{z}_{m}^{(b_{0})}$
are the real channel and the channel noise, respectively, modeled as Gaussian random matrix with zero
mean and variance~$1/\emph{N}$. The channel estimate error of MUE $m$ is denoted by $\tau _m$; in case of perfect CSI, $\tau _m = 0$. Similarly, let $\mathbf{H}^{(b_{0}), \text{S}} \in \mathbb{C}^{\emph{N} \times \emph{S}}$ and $\mathbf{H}^{(b_{0}), \text{C}} \in
\mathbb{C}^{\emph{N} \times \emph{S}}$ denote the channel matrices from the MBS antennas to SCs and SUEs,
respectively. Let ${h}_{i}^{ {(b_s)}}$ denote the channel propagation from SC $b_s$ to any receiver $i$. Let $c_s$ denote the SUE served by the SC $b_s$.
\section{ Joint Access/Backhaul Scheduling and Interference Mitigation Problem Formulation}
\label{Op-Form}
We address the problem of DL scheduling at the MBS to simultaneously provide data transmission to MUEs and wireless backhaul to SCs. We define the scheduling vector $\mathbf{\us}(t) = (\us_{1}(t), \us_{2}(t), \ldots, \us_{\emph{K}}(t))$ to determine the subset of users served at time slot $t$, where $\us_{k}(t) = 1$ means user $k$ is served at time slot $t$ and $\us_{k}(t) = 0$ otherwise.
\subsection{Downlink Transmission Signal}
\label{DL-Signal}
The MBS serves two types of users: MUEs and SCs, let $p_{m}^{(b_0)}$, $p_{s}^{(b_0)}$, and $P^{(b_0)}$ denote the DL MBS transmit power assigned to MUE $m$, SC $b_{s}$, and the maximum transmit power at the MBS, respectively. Let $\pc_{\sue_{s}}^{(b_s)}$ denote the DL transmit power of SC $b_{s}$ assigned to SUE ${\sue_{s}}$. Although SC exploits FD capability to double capacity, SC causes unwanted FD interference including cross-tier interference to adjacent MUEs (or other SCs) and co-tier interference to other SUEs. Moreover, inspired by~\cite{wagner2012l}, we invoke RMT to get closed-form expression for the user data rate as $\emph{N}, \emph{K} \gg 1$. In order to convert the interference channel to multiple-input single-output (MISO) channel, we design a precoder at MBS and propose an operation mode policy to control FD interference due to the SC DL transmissions such that the total FD interference at receiver is treated as noise.

\begin{definition}\label{USP}[Operation Mode Policy] We define $\boldsymbol{\op}$ as the operation mode to control the SC transmission to reduce FD interference. The operation mode is expressed as $\boldsymbol{\op}(t) = \{ \op^{(b_s)}(t)~|~ \op^{(b_s)}(t) \in \{0, 1\}, \forall s \in \mathcal{S}  \}$. If SC $b_{s}$ operates in FD mode, then $\op^{(b_s)}(t) = 1$. When SC $b_{s}$ operates in half-duplex (HD) mode, i.e., $\op^{(b_s)}(t) = 0$.
\end{definition}

We assume that the MBS uses a precoding scheme, $\mathbf{V} = [\mathbf{v}_1, \mathbf{v}_2, \ldots, \mathbf{v}_{\emph{K}}] \in \mathbb{C}^{\emph{N} \times \emph{K}}$. We apply the technique in~\cite{Liu2014} to design the precoder $\mathbf{V} = \mathbf{U} \mathbf{T}$, where $\mathbf{U} \in \mathbb{C}^{\emph{N} \times \emph{N}_{\mathrm{i}}}$ and  $\mathbf{T} \in \mathbb{C}^{\emph{N}_{\mathrm{i}} \times \emph{K}}$ are used to control cross-tier and co-tier interference, respectively. Here, $\emph{N}_{\mathrm{i}}$ antennas are used to mitigate the cross-tier interference with $\emph{N}_{\mathrm{i}} < \emph{N}$, where the subscript $\mathrm{i}$ stands for ``interference". The precoder $\mathbf{U}$ is chosen such that
\begin{equation}\label{zero-ICI}
\mathbf{U}^{\dag} \textstyle \sum _{s = 1}^{\emph{S}} \op^{(b_s)} \mathbf{\Theta}_{\sue_{s}}^{(b_0)} = 0,
\end{equation}
where $\mathbf{\Theta}_{\sue_{s}}^{(b_0)} \in \mathbb{C}^{\emph{N} \times \emph{N}}$ is the correlation matrix between MBS antennas and SUE ${\sue_{s}}$. Note that $\op^{(b_s)}$ determines that SUE ${\sue_{s}}$ is served or not, and $\mathbf{U}^{\dag}$ denotes the Hermitian transpose of matrix $\mathbf{U}$. The precoder $\mathbf{T}$ is designed to adapt to the real time CSI based on $\mathbf{\hat{H}^{\dag} \mathbf{U} } \in \mathbb{C}^{\emph{K} \times \emph{N}_{\mathrm{i}}}$, where $\mathbf{\hat{H}} = [ \mathbf{\hat{h}}^{(b_{0})}]_{k \in \mathcal{K}}^{\dag}$. In this paper, we consider the regularized zero-forcing (RZF) precoding\footnote{Other precoders are left for future work.} given by $\mathbf{T} = \big(\mathbf{U}^{\dag} \mathbf{\hat{H}}^{\dag}  \mathbf{\hat{H}}\mathbf{U} + \emph{N} \alpha \mathbf{I}_{\emph{N}_{\mathrm{i}}}\big)^{-1}  \mathbf{U}^{\dag} \mathbf{\hat{H}}^{\dag}$, where the regularization parameter $\alpha > 0$ is scaled by $\emph{N}$ to ensure that the matrix $\mathbf{U}^{\dag} \mathbf{\hat{H}}^{\dag} \mathbf{\hat{H}}\mathbf{U} + \emph{N} \alpha \mathbf{I}_{N_\mathrm{i}}$ is well conditioned as $\emph{N} \rightarrow \infty$. The precoder $\mathbf{T}$ is chosen to satisfy the power constraint $\text{Tr}\big (\mathbf{P} \mathbf{T}^{\dag} \mathbf{T} \big) \leq P^{(b_0)}$, where $\mathbf{P} = \text{diag}(p_1^{(b_0)}, p_2^{(b_0)}, \ldots, p_{\emph{K}}^{(b_0)})$. By utilizing a massive number of antennas at MBS, a large spatial degree of freedom is partly utilized to serve MUEs and FD-enabled SCs, while the remaining degrees of freedom are used to mitigate cross-tier interference. In massive MIMO system, the total number of antennas is considered as the degree of freedom~\cite{marzt2010non}. Hence, we have the antenna constraint for scheduling and operation mode such that $\sum_{k = 1}^{\emph{K}} \us_{k}(t) + \sum_{s =  1}^{\emph{S}} \op^{(b_{s})}(t) \leq \emph{N}$. For notational simplicity, we remove the time dependency from the symbols throughout the discussion. The received signal $y_{m}^{(b_0)}$  at each MUE $m \in \mathcal{M}$ at time instant $t$ is given by

\begin{equation} \label{SINR-MUE-1}
\begin{split}
\displaystyle y_{m}^{(b_0)} =~& \us_{m}\sqrt{p_{m}^{(b_0)}} \mathbf{h}_{m}^{(b_0) \dag} \mathbf{v}_{m} x_{m}^{(b_0)}
+ \underbrace{\textstyle \sum _{s = 1}^{\emph{S}} \op^{(b_s)} \sqrt{\pc_{\sue_s}^{(b_s)}} h_{m}^{(b_s) \dag} {x}_{\sue_s}^{(b_s)}}_{\text{cross-tier interference}}
\\& +  \underbrace{ \textstyle\sum _{k = 1, k \neq m}^{\emph{K} } \us_{k}\sqrt{p_{k}^{(b_0)}} \mathbf{{h}}_{m}^{(b_0) \dag} \mathbf{v}_{k} x_{k}^{(b_0)}}_{\text{co-tier interference}} + \eta_{m},
\end{split}
\end{equation}
where $x_{m}^{(b_0)}$ is the signal symbol from the MBS to the MUE $m$, $\eta_{m} \sim \mathcal{CN}(0,1)$ is the thermal noise at MUE $m$, and $\mathbf{v}_{m}$ is the precoding vectors of MUE $m$. ${x}_{\sue_s}^{{(b_s)}}$ is the transmit signal symbol from SC $b_{s}$ to SUE $\sue_s$.

At time instant $t$, the received signal $y_{s}^{(b_0)}$ at each SC $b_{s} \in \mathcal{S}$ suffers from self-interference, cross-tier interference, and co-tier interference which is given by

\begin{equation} \label{SINR-SC-1}
\begin{split}
y_{s}^{(b_0)}  =~& \us_{s} \sqrt{p_{s}^{(b_0)}}  \mathbf{h}_{s}^{(b_0) \dag} \mathbf{v}_{s}^{\text{}} x_{s}^{(b_0)}
+ \underbrace{\op^{(b_s)} \sqrt{\pc_{\sue_s}^{(b_s)}} {h}_{s}^{ {(b_s)} \dag} {x}_{\sue_s}^{ (b_s) }}_{\text{self-interference}}
\\& + \underbrace{\textstyle \sum _{s' = 1, s' \neq s}^{\emph{S}} \op^{(b_{s'})}\sqrt{\pc_{\sue_{s'}}^{(b_{s'})}} {h}_{s}^{ (b_{s'}) \dag} {x}_{\sue_{s'}}^{ (b_{s'}) }}_{\text{cross-tier interference}}
\\& + \underbrace{\textstyle \sum _{k = 1, k \neq s}^{\emph{K} } \us_{k} \sqrt{p_{k}^{(b_0)}} \mathbf{h}_{s}^{ (b_0) \dag} \mathbf{v}_{k} x_{k}^{ (b_0)} }_{\text{co-tier interference}} + \eta_{s},
\end{split}
\end{equation}
where $\eta_{s} \sim \mathcal{CN}(0,1)$ is the thermal noise of the SC $b_{s}$. In this work, the FD capability is leveraged at SCs.

The received signal at SUE $\sue_s$ from its serving SC $b_{s}$ is interfered by the DL signals from other SCs and the MBS. At time instant $t$, it is given by

\begin{equation} \label{SINR-Sue-1}
\begin{split}
y_{\sue_s}^{(b_s)} =~& \op^{(b_{s})} \sqrt{\pc_{\sue_s}^{(b_s)}} {h}_{\sue_s}^{(b_s) \dag } x_{\sue_s}^{(b_s)}
+ \underbrace{ \textstyle \sum _{k = 1}^{\emph{K} } \us_{k} \sqrt{p_{k}^{(b_{0})}} \mathbf{h}_{\sue_s}^{ (b_{0}) \dag} \mathbf{v}_{k} x_{k}^{(b_{0})} }_{\text{cross-tier interference}}
\\& + \underbrace{\textstyle \sum _{s' = 1, s' \neq s}^{\emph{S} } \op^{(b_{s'})} \sqrt{\pc_{\sue_{s'}}^{(b_{s'})}} {h}_{\sue_s}^{(b_{s'}) \dag} x_{\sue_{s'}}^{(b_{s'})} }_{\text{co-tier interference}} + \eta_{\sue_s},
\end{split}
\end{equation}
where $x_{\sue_s}^{(b_s)}$ is the transmit data symbol from the SC $b_{s}$ to its own SUE $\sue_s$ and $\eta_{\sue_s} \sim \mathcal{CN}(0,1)$ is the SUE thermal noise.

According to~(\ref{zero-ICI})-(\ref{SINR-Sue-1}), the $\mathrm{SINRs}$ of  MUE $m \in \mathcal{M}$, SC $b_{s} \in \mathcal{S}$, and SUE ${\sue_{s}}$ are given by:
\begin{equation} \label{SINR-MUE-2}
\begin{split} \gamma_m^{(b_0)} =  \frac { \us_{m} p_{m}^{(b_0)} |\mathbf{h}_{m}^{(b_0) \dag} \mathbf{v}_{m} |^2 }  {  {\textstyle \sum \limits_{k \neq m} \us_{k} p_{k}^{(b_0)} | \mathbf{h}_{m}^{(b_0) \dag} \mathbf{v}_{k}|^2  + \textstyle \sum \limits_{s} \op^{(b_{s})} \pc_{\sue_{s}}^{(b_s)} |h_{m}^{(b_s) \dag}|^2 + 1}}  \end{split}
\end{equation}
\begin{equation} \label{SINR-SC-2}
\begin{split}
\gamma_{s}^{(b_0)} =  \frac { \us_{s} p_{s}^{(b_0)} | \mathbf{h}_{s}^{(b_0) \dag} \mathbf{v}_{s}|^2 }{  {\textstyle \sum \limits_{k \neq
s} \us_{k} p_{k}^{(b_0)} |\mathbf{h}_{s}^{(b_0) \dag} \mathbf{v}_{k}|^2  + \textstyle \sum \limits_{s' \neq s} \op^{(b_{s'})} \pc_{\sue_{s'}}^{(b_{s'})} |{h}_{s}^{(b_{s'}) \dag}|^2 + 1 }},
\end{split}
\end{equation}
\begin{equation} \label{SINR-SUE-2}
\begin{split}
\gamma_{\sue_{s}}^{(b_s)} =   \frac {\op^{(b_{s})} \pc_c^{(b_s)} |{h}_{\sue_{s}}^{(b_s) \dag}|^2} { {\textstyle \sum \limits_{s' \neq s} \op^{(b_{\sue_{s'}})} \pc_{\sue_{s'}}^{(b_{s'})} |{h}_{\sue_{s}}^{ {(b_{s'}) } \dag}|^2  + 1}}.
\end{split}
\end{equation}
\subsection{Downlink Scheduling Problem}
\label{NUM-Pro}
 We consider a joint optimization of scheduling $\mathbf{l}$, operation mode $\boldsymbol{\op}$, interference mitigation $\mathbf{U}$, and transmit power allocation $\mathbf{p} = (p_{1}^{(b_0)}, p_{2}^{(b_0)}, \ldots, p_{\emph{K}}^{(b_0)})$ that satisfies the transmit power budget of MBS, i.e., $\text{Tr}\big (\mathbf{P} \mathbf{T}^{\dag} \mathbf{T} \big) \leq P^{(b_0)}$. We define $\EP_{i}^{(b_s)} = \frac{p_{c_s}^{(b_s)} |h_{i}^{(b_s) \dag}|^2}{|\eta_i|^2}$ and $\epsilon_{o}$ as the FD interference to noise ratio ($\mathrm{INR}$) from SC $b_{s}$ to any scheduled receiver $i$, and the allowed FD $\mathrm{INR}$ threshold, respectively. The FD $\mathrm{INR}$ threshold is defined such that $\textstyle \sum_{ i = 1 }^{\emph{K}} \textstyle \sum_{s = 1}^{ \emph{S} } \EP_{i}^{(b_s) } \leq \epsilon_{o}$ in which the total FD interference is considered as noise. Under the operation mode policy, we schedule the receiver $i$ and enable the transmission of SC $b_s$ as long as $\textstyle \sum_{i = 1}^{\emph{K}} \textstyle \sum_{s = 1}^{\emph{S}} \us_{i} \op^{(b_s)} \EP_{i}^{(b_s)} \leq \epsilon_{o}$.

 Let $\mathbf{\Lambda}^{o} = \{ \mathbf{\us}, \boldsymbol{\op}\}$ be a composite control variable of scheduling and operation mode. We define $\mathbf{\Lambda} = \{ \mathbf{\Lambda}^{o}, \mathbf{U}, \mathbf{p}\}$ as a composite control variable, which adapts to the spatial channel correlation matrix $\mathbf{\Theta}$. For a given $\mathbf{\Lambda}$ that satisfies~(\ref{zero-ICI}) and operation mode policy, the respective Ergodic data rates of MUE, SC, and SUE are:
 \begin{equation} \label{Rt01}
\begin{split}
r_{m}(\mathbf{\Lambda}|\mathbf{\Theta})  =  \mathbb{E} \big[  \log \big( 1 + \gamma_m^{(b_0)} \big)\big],  \nonumber \\
r_{s}(\mathbf{\Lambda}|\mathbf{\Theta})  = \mathbb{E} \big[ \log \big( 1 +  \gamma_s^{(b_0)}  \big) \big],  \text{and}\nonumber \\
r_{\sue_{s}}(\mathbf{\Lambda}|\mathbf{\Theta})  =  \mathbb{E} \big[ \log  \big( 1 + \gamma_{\sue_{s}}^{(b_s)}  \big) \big]. \nonumber
\end{split}
\end{equation}

\begin{definition}  For any vector $\mathbf{x}(t) = ({x}_{1}(t), ...,{x}_{\emph{K}}(t))$, let $\bar{\mathbf{x}} = (\bar{x}_{1}, \cdots,\bar{x}_{\emph{K}})$ denote the time average expectation of $\mathbf{x}(t)$, where $\textstyle \bar{\mathbf{x}} \triangleq \lim\limits_{t \to \infty} \frac{1}{t}  \sum\limits_{\tau=0}^{t-1} \mathbb{E} [\mathbf{x}(\tau)] $. Similarly, $\textstyle \bar{\mathbf{r}} \triangleq \lim\limits_{t \to \infty} \frac{1}{t}  \sum\limits_{\tau=0}^{t-1} \mathbb{E}[\mathbf{r}(\tau)]$ denotes the time average expectation of the Ergodic data rate.
\end{definition}

For a given composite control variable $\mathbf{\Lambda}$ that adapts to the spatial channel correlation matrix $\mathbf{\Theta}$, the average data rate region is defined as the convex hull of the average data rate of users, which is expressed as:

\begin{equation}\label{eq:rate-region}
\begin{split}
\textstyle \mathcal{R} \triangleq & \big \{ \bar{\mathbf{r}}(\mathbf{\Lambda}|\mathbf{\Theta}) \in \mathbf{R}^{\emph{K}}_{+}~|~\mathbf{\us} \in \{0, 1\}^{\emph{K}},  \mathbf{\op} \in \{0, 1\}^{\emph{S}}, \\
\begin{split}
\end{split}
& \textstyle\sum_{k = 1}^{\emph{K}} \us_{k} + \textstyle \sum_{s =  1}^{\emph{S}} \op^{(b_{s})} \leq \emph{N}, \\
& \textstyle \sum _{ i = 1 }^{\emph{K}} \textstyle \sum_{s = 1}^{\emph{S}} \us_{k}\op^{(b_s)} \EP_{k}^{(b_s)} \leq \epsilon_{o},\\
&\text{Tr}\big (\mathbf{P} \mathbf{T}^{\dag} \mathbf{T} \big) \leq P^{(b_0)}, \\
& \mathbf{U}^{\dag} \textstyle \sum _{s = 1}^{\emph{S}} \op^{(b_{s})} \mathbf{\Theta}_{c_s}^{(b_0)} = 0  \big \},\nonumber
\end{split}
\end{equation}
where $\bar{\mathbf{r}}(\mathbf{\Lambda}|\mathbf{\Theta}) = (\bar{r}_{1}(\mathbf{\Lambda}|\mathbf{\Theta}), \ldots, \bar{r}_{\emph{K}}(\mathbf{\Lambda}|\mathbf{\Theta}))^{T}$. Following the results from~\cite{pareto2011}, the boundary points of the rate regime with total power constraint and no self-interference are Pareto-optimal\footnote{The Pareto optimal is the set of user rates at which it is impossible to improve any of the rates without simultaneously decreasing at least one of the others.}. Moreover, according to~\cite[Proposition 1]{pareto2012}, if the $\mathrm{INR}$ covariance matrices approach the identity matrix, the Pareto rate regime of the MIMO interference system is convex. Hence, the rate regime is Pareto-optimal, and thus, convex under above constraints.

Let us assume that SCs act as relays to forward data to the SUEs. If the MBS transmits data to SC $b_s$, but the transmission of SC $b_s$ is disabled, it cannot serve its SUE. Hence, we define $\mathbf{D(t)} = (D_1(t), D_2(t), \ldots, D_{\emph{S}}(t))$ as a data queue at SCs, where at each time slot $t$, the wireless backhaul queue at SC $b_s$ is
\begin{equation}\label{queueD}
{D}_{s}(t + 1) = \mathrm{max}[{D}_{s}(t) + {r}_{s}(t) - {r}_{\sue_{s}}(t), 0].
\end{equation}
We define the network utility function $f_{0}$ to be non-decreasing, concave over the convex region $\mathcal{R}$ for a given $\mathbf{\Theta}$. The objective is to maximize the network utility under wireless backhaul queue constraints and imperfect CSI. Thus, the network utility maximization (NUM) problem is given by,
\begin{subequations}\label{eq:Obj-Formulate-0}
\begin{align}
& \max_{\bar{\mathbf{r}}}
& & f_{0} (\bar{\mathbf{r}}) \label{eq-o1}\\
& \text{subject to}
&&  \bar{\mathbf{r}} \in \mathcal{R}, \label{eq-o2} \\
&&& \bar{\mathbf{D}} < \infty.\label{eq-o3}
\end{align}
\end{subequations}
where $f_{0} ( \bar{\mathbf{r}} ) =   \textstyle \sum _{k = 1}^{\emph{K}} \omega_{k}(t) f( \bar{r}_{k} )$, and $\omega_{k}(t) \geq 0$ is the weight of user $k$, $f(\cdot)$ is assumed to be twice differentiable, concave, and increasing L-Lipschitz function for all $\bar{r} \geq 0$. Solving~(\ref{eq:Obj-Formulate-0}) is non-trivial since the average rate region $\mathcal{R}$ does not have a tractable form. To overcome this challenge, we need to find closed-form expressions of the data rate and the average transmit power.
\subsection{Closed-Form Expression via Deterministic Equivalent}
\label{Closed-Form}
We invoke recent results from RMT in order to get the deterministic equivalent. As $K, N \rightarrow \infty$, by applying the techniques in~\cite[Theorem
2]{wagner2012l}, for small $\alpha$ the deterministic equivalent of the asymptotic $\mathrm{SINR}$s of UEs (\ref{SINR-MUE-2}-\ref{SINR-SC-2}) is
\begin{equation}\label{SINR-MUE-3}
\gamma_{m}^{(b_0)}(\mathbf{\Lambda}|\mathbf{\Theta}) \xrightarrow{a.s.}  \frac{\us_m p_{m}^{(b_0) } (1 - \tau_m^2)}{1 + \textstyle \sum _{s = 1}^{\emph{S}} \op^{(b_s)} \EP_{m}^{(b_s)}}, \nonumber
\end{equation}
\begin{equation}\label{SINR-SC-3}
\gamma_{s}^{(b_0)}(\mathbf{\Lambda}|\mathbf{\Theta}) \xrightarrow{a.s.}  \frac{\us_s p_{s}^{(b_0)}}{1 + \textstyle \sum _{s' = 1, s'\neq s}^{\emph{S}} \op^{(b_{s'})} \EP_{s}^{(b_{s'})}},\nonumber
\end{equation}
where $\xrightarrow{a.s.}$ denotes the almost sure convergence, please see the proof in Appendix~\ref{Appendix1}. The precoder $\mathbf{T}$ is designed to satisfy a total transmit power constraint and takes into account the scheduling $\mathbf{l}$ and $\mathbf{U}$, and thus, we obtain $\textstyle \frac{1}{\emph{N}} \textstyle \sum _{k = 1}^{\emph{K}} \frac{ {p}_{k}^{(b_0)}}{  \Omega_{k} } - P^{(b_0)} \leq 0$. Here, $\Omega_{k} = \textstyle \frac{1}{\emph{N}} \mathrm{Tr} (\mathbf{\tilde{\Theta}}_k \mathbf{G})$ forms the unique positive solution of which is the Stieltjes transform of nonnegative finite measure~\cite[Theorem 1]{wagner2012l}, where  $\textstyle \mathbf{G} =  \big( \frac{1}{\emph{N}} \sum _{k = 1}^{\emph{K}} \frac{ \mathbf{\tilde{\Theta}}_k }{ \alpha + \Omega_{k}} + \mathbf{I}_{\emph{N}_\mathrm{i}} \big)^{-1}$ and $\mathbf{\tilde{\Theta}}_k = \mathbf{U} \mathbf{U}^{\dag} \mathbf{\Theta}_k^{(b_0)} \mathbf{U} \mathbf{U}^{\dag}$.

Although the closed-form expression of time average of data rate and transmit power is obtained, our problem considers a time-average optimization with a large number of control variables and dynamic traffic load over the convex region for a given composite control variable $\mathbf{\Lambda}$ and $\mathbf{\Theta}$. Moreover, the goal is to maximize the aggregate network utility subject to queue stability in which the Lyapunov optimization framework can be utilized effectively, by means of drift-plus-penalty technique~\cite{neely2010S} to solve the joint scheduling and interference mitigation problem.
\section{ Lyapunov Optimization Framework}
\label{LOF}
The network operation is modeled as a queueing network that operates in discrete time $t \in \{0, 1, 2, \dots \}$. Let $a_k(t)$ denote the bursty data arrival destined for each user $k$, $\text{i.i.d}$ over time slot $t$. Let $\mathbf{Q}(t)$ denote the vector of transmission queue blacklogs at MBS at slot $t$. The evolution of $\mathbf{Q}(t)$ is
\begin{equation}\label{queueQ}
{Q}_{k}(t + 1) = \mathrm{max}[{Q}_{k}(t) - {r}_{k}(t), 0] + {a}_{k}(t), \forall k \in \mathcal{K}.
\end{equation}
Here, we define the bound of the traffic arrival of user $k$ such that $0 \leq {a}_{k}(t) \leq a_{k}^{\mathrm{max}}$, for some constant $a_{k}^{{\max}} < \infty$. Here, we define the upper bound of traffic data $r_k^{\mathrm{max}}(t) < \infty$ for user $k$ at time slot $t$. The set constraint~(\ref{eq-o2}) is replaced by an inequivalent set constraint by introducing $\text{auxiliary variables}$ $\boldsymbol{\varphi}(t) \in \mathcal{R}$, $\boldsymbol{\varphi}(t) = \big(\varphi_{1}(t), \ldots, \varphi_{K}(t) \big)$ that satisfies $\bar{\varphi}_{k} \leq \bar{r}_{k}$, where $\bar{\varphi}_{k} \triangleq \lim\limits_{t \to \infty} \frac{1}{t} \textstyle \sum\limits_{\tau=0}^{t-1} \mathbb{E} \big[ \varphi_{k}(\tau))  \big] $. The wireless backhaul queue can be rewritten as
\begin{equation}\label{queueD}
{D}_{s}(t + 1) = {\max}[{D}_{s}(t) + \varphi_{s}(t) - {r}_{\sue_{s}}(t), 0].
\end{equation}
For a given $\mathbf{\Lambda}$ and $\mathbf{\Theta}$, the optimization problem~(\ref{eq:Obj-Formulate-0}) with the network stability can be posed as
\begin{subequations}\label{eq:Obj-Formulate-1}
\begin{align}
& \min_{ \bar{\boldsymbol{\varphi}} }
&& - f_{0}( \bar{\boldsymbol{\varphi}}) \label{eq-1a} \\
& \text{subject to}
&&  \bar{\varphi}_{k} - \bar{r}_{k} \leq 0, \forall k \in \mathcal{K},  \label{eq-1b} \\
&&& \bar{\mathbf{D}} < \infty, \mathbf{\bar{Q}} < \infty. \label{eq-1d}
\end{align}
\end{subequations}
In order to ensure the inequality constraint~(\ref{eq-1b}), we introduce a virtual queue vector ${Y}(t)$ where
\begin{equation}\label{queueY}
{Y}_{k}(t + 1) = \mathrm{max}[{Y}_{k}(t) + \varphi_{k}(t) - {r}_{k}(t), 0].
\end{equation}
We define the queue backlog vector as $\mathbf{\Sigma}(t) = \big[ \mathbf{Q}(t), \mathbf{Y}(t), \mathbf{D}(t)\big]$ which involves all constraints of~(\ref{eq:Obj-Formulate-1}). The Lyapunov function can be written as ${L}(\mathbf{\Sigma}(t)) \triangleq \frac{1}{2} \big[ \textstyle \sum \limits_{k = 1}^{\emph{K}} {Q}_{k}(t)^{2} + \textstyle \sum \limits_{k = 1}^{\emph{K}} {Y}_{k}(t)^{2} + \textstyle \sum \limits_{s = 1}^{\emph{S}} {D}_{s}(t)^{2}\big]$. For each time slot $t$, $\mathbf{\Delta}(\mathbf{\Sigma}(t))$ denotes the Lyapunov drift given by $\mathbf{\Delta}(\mathbf{\Sigma}(t)) \triangleq \mathbb{E} \big [ {L}(\mathbf{\Sigma}(t+1))  - {L}(\mathbf{\Sigma}(t))| \mathbf{\Sigma}(t) \big]$.

\begin{figure}[!h]
\begin{minipage}[t]{1\linewidth}
    \centering
    \includegraphics[scale=0.6]{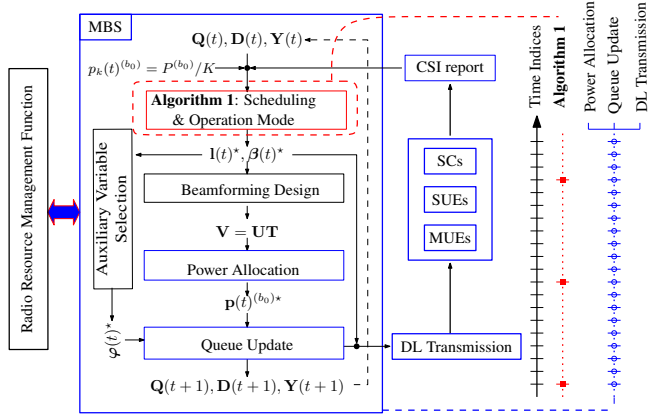}
    \caption{Joint Scheduling and Interference Mitigation Algorithm.}\label{Algorithm0}
\end{minipage}
\end{figure}

Noting that $\mathrm{max}[a,0]^2 \leq  a^2$ and $(a \pm b)^2 \leq a^2 \pm 2ab + b^2$ for any real positive number $a, b$, and thus, by neglecting the index $t$ we have, $(\mathrm{max}[{Q}_{k} - r_{k}, 0] + {a}_{k})^{2} - {Q}_{k}^2 \leq 2 {Q}_{k}(a_{k}- {r}_{k}) + (a_{k}- {r}_{k})^{2}$, $\mathrm{max}[{Y}_{k} + \varphi_{k}- {r}_{k}, 0]^{2} - {Y}_{k}^2 \leq 2 {Y}_{k}(\varphi_{k}- {r}_{k}) + (\varphi_{k}- {r}_{k})^{2}$, $\mathrm{max}[{D}_{s} + \varphi_{s}- {r}_{\sue_{s}}, 0]^{2} - {D}_{s}^2 \leq 2 {D}_{s}(\varphi_{s}- {r}_{\sue_{s}}) + (\varphi_{s}- {r}_{\sue_{s}})^{2}$. Assume that $\boldsymbol{\varphi}_k \in \mathcal{R}$ and a feasible $\mathbf{\us}$ for all $t$ and all possible $\mathbf{\Sigma}(t)$, we have
\begin{equation}\label{eq:Lya-Driff-Pen}
\begin{aligned}
\mathbf{\Delta}(\mathbf{\Sigma}(t)) &\leq  \Psi +  \textstyle \sum_{k = 1}^{\emph{K}} {Q}_{k}(t) \mathbb{E} \big[  {a}_{k}(t) - {r}_{k}(t) | \mathbf{\Sigma}(t) \big ]
\\&+ \textstyle \sum _{k = 1}^{\emph{K}} {Y}_{k}(t) \mathbb{E} \big[ \varphi_{k}(t) - r_{k}(t) | \mathbf{\Sigma}(t) \big]
\\&+ \textstyle \sum _{s = 1}^{\emph{S}} {D}_{s}(t)\mathbb{E} \big[  {\varphi}_{s}(t)  - {r}_{\sue_{s}}(t) |\mathbf{\Sigma}(t) \big ].
\end{aligned}
\end{equation}
Here $\mathbf{\Delta}(\mathbf{\Sigma}(t)) \leq  \Pi$, where  $\Pi$ represents the $\text{R.H.S}$ of~(\ref{eq:Lya-Driff-Pen}), and $\Psi$ is a finite constant that satisfies $\Psi \geq      \frac{1}{2} \sum_{k = 1}^{\emph{K}}\mathbb{E} \big[  \big({a}_{k}(t) - {r}_{k}(t)\big)^2 | \mathbf{\Sigma}(t) \big ] +   \frac{1}{2} \sum_{k = 1}^{\emph{K}}  \mathbb{E} \big[  \big(\varphi_{k}(t) - {r}_{k}(t)\big)^2 | \mathbf{\Sigma}(t) \big ] +   \frac{1}{2} \sum_{s = 1}^{\emph{S}}\mathbb{E} \big[  \big( {\varphi}_{s}(t) - {r}_{\sue_{s}}(t) \big)^2 | \mathbf{\Sigma}(t)\big ]$, for all $t$ and all possible $\mathbf{\Sigma}(t)$.
We apply the Lyapunov drift-plus-penalty technique, where the solution of~(\ref{eq:Obj-Formulate-1}) is obtained by minimizing the Lyapunov drift and the objective function, i.e.,  $\text{min}~\Pi-\nu \mathbb{E}[f_{0} ( \boldsymbol {\varphi}(t) )]$, where the parameter $\nu$ is chosen as a non-negative constant to control optimality and queue backlogs. Since $\Psi$ is finite, the problem becomes minimizing~(\ref{FinalEQ}), which is decoupled over scheduling and operation mode variables ($1\star$), auxiliary variables ($2\star$), and precoder and power allocation variables ($3\star$), respectively. Hence, the respective variables can be found independently by minimizing the individual term at each time. Fig.~\ref{Algorithm0} summarizes the relationship among subproblems.

\begin{floatEq}
\begin{equation}\label{FinalEQ}
\Big[ \big[\overbrace{-\textstyle \sum _{k} \big ( {Q}_{k}(t) + {Y}_{k}(t) \big) {r}_{k}(\mathbf{\Lambda}(t)) }^{\text{Impact of network queue, virtual queue, and } \mathbf{\Lambda}} \big]_{3\star} \overbrace{-\textstyle \sum _{s} {D}_{s}(t) {r}_{\sue_{s}}(\op^{(b_s)}(t))}^{\text{Impact of SC queue and } \boldsymbol{\op}}\Big]_{1\star}  + \Big[ \overbrace{\textstyle \sum _{k} {Y}_{k}(t) \varphi_{k}(t) +  \textstyle \sum _{s} {D}_{s}(t) {\varphi}_{s}(t)}^{\text{Impact of virtual queue, SC queue, and auxiliaries}}  \overbrace{- \nu f_{0} ( \boldsymbol {\varphi}(t) )}^{\text{penalty}}\Big]_{2\star}.
\end{equation}
\end{floatEq}

\subsection{Step 1: Joint In-Band Scheduling and FD or HD Operation Mode Control}
\label{user-scheduling}
The joint in-band scheduling and SC operation mode control problem is a mixed-integer program with a non-convex objective function and non-convex interference constraints. To solve this problem, we first relax the integer constraints to linear constraints as continuous values. Secondly, to deal with non-convex interference constraint, we apply the SCA method~\cite[Lemma 3.5]{beck2010seq}, where at each iteration $i$ the non-convex constraints are approximated by upper convex approximations. Hence, we solve iteratively the problem  by updating the variables until convergence to obtain a Karush-Kuhn-Tucker (KKT) point. The interference constraint is replaced by $\sum_{i=1}^{\emph{K}} \sum_{s =1}^{\emph{S}} \big( \frac{\lambda_{ks}^{(i)} \us^{2}_{k}(t)}{2} + \frac{(\op^{(b_s)})^{2}(t)}{2 \lambda_{ks}^{(i)}} \big) \EP_{k}^{(b_s)}(t) - \epsilon_{o} \leq 0$, for every fixed positive value $\lambda_{ks}^{(i)}$. Finally, instead of minimizing the non-convex objective function $(2\star)$, we minimize its upper bound by replacing the denominators, i.e., $1 + \textstyle \sum _{s = 1}^{\emph{S}} \op^{(b_s)} \EP_{m}^{(b_s)}$ with the largest bound, i.e., $1 + \epsilon_{0}$. This upper bound is obtained due to interference constraint $\sum_{i=1}^{\emph{K}} \sum_{s =1}^{\emph{S}} \us_i \op^{(b_s)} \EP_{i}^{(b_s)} \leq  \epsilon_{o}$ and we assume the feasibility of problem exists such that the total FD interference at each receiver is less than the total FD interference threshold $\epsilon_{0}$. Hence, we obtain the upper bound as below and the optimal value of $\mathbf{\Lambda}^{o}$ is given by

\begin{subequations}\label{Optimal-US}
\begin{align}
& \min_{\mathbf{l},\mathbf{\op}} \nonumber
&&- \textstyle \sum _{s = 1}^{\emph{S}} {D}_{s}(t) \log  \big( 1 +  \frac{\op^{(b_s)}(t) p_{\sue_{s}}^{(b_s)} |{h}_{\sue_{s}}^{(b_s) \dag}|^2}{1 + \epsilon_{0}  } \big) \nonumber\\
&&& - \textstyle \sum _{k = 1}^{\emph{K}} A_{k}(t) \log \big( 1 +  \frac{\us_k(t) p_{k}^{(b_0) } (1 - \tau_k^2)} {1 + \epsilon_{0} } \big) \label{eq-us1}\\
& \text{subject to}
&&    \us_k(t), \op^{(b_s)}(t) \in [0, 1], \forall k \in \mathcal{K}, \forall s \in \mathcal{S},\label{eq-us2}\\
&&&   \textstyle \sum_{k = 1}^{\emph{K}} \us_{k}(t) + \sum_{s = 1}^{\emph{S}} \op^{(b_s)}(t) \leq \emph{N},\label{eq-us3}\\
&&&   \textstyle \sum\limits_{i = 1}^{\emph{K}} \textstyle \sum\limits_{s = 1}^{\emph{S}} \big( \frac{\lambda_{ks}^{(i)} \us^{2}_{k}(t)}{2} + \frac{(\op^{(b_s)})^{2}(t)}{2 \lambda_{ks}^{(i)}} \big) \EP_{k}^{(b_s)}(t) \leq  \epsilon_{o}, \label{eq-us5}\\
&&&   r_k(\us_k(t),\op^{(b_s)}(t)) \in \mathcal{R}\label{eq-us4},
\end{align}
\end{subequations}
where $A_{k}(t) = {Q}_{k}(t) + {Y}_{k}(t)$. At each time slot $t$, the joint in-band scheduling and SC operation mode control problem is outlined as Algorithm~\ref{algUS}. We numerically observe that the SCA-based Algorithm~\ref{algUS} converges quickly after few iterations and yields a solution of many scheduling and operation variables close or equal to binary. Hence, we apply a binary search algorithm in order to obtain
a low-complexity search algorithm to convert the continuous relaxation solution to the integer solution~\cite{li2014e}. To ensure that all users will be served, each user is set to have the same transmit power to find the best scheduled users. Moreover, the scheduling will be performed in a long-term period, while the power allocation problem is executed in a short-term period.

\begin{algorithm}                   
\caption{Joint in-band scheduling and SC operation mode control problem}
\label{algUS}
\begin{algorithmic}                    
    \STATE Initialization $i = 0$, $\lambda_{ks}^{(i)} = \text{randomly positive}$.
    \REPEAT
        \STATE $\text{Solve}$~(\ref{Optimal-US}) with $\lambda_{ks}^{(i)}$ to get optimal value $\mathbf{\Lambda}^{o\star}$.
        \STATE $\text{Update}$ $\mathbf{\Lambda}^{o(i)} := \mathbf{\Lambda}^{o\star}$ and $\lambda_{ks}^{(i+1)} := \frac{\op^{(b_s)(i)}}{\us^{(i)}_{k}}$; $i := i + 1$.
    \UNTIL{\text{Convergence}}
\end{algorithmic}
\end{algorithm}

\subsection{Step 2: Selection of Auxiliary Variable}
\label{AV}
The optimal auxiliary variables are computed by

\begin{subequations}\label{eq:Optimal-AR}
\begin{align}
& \min_{\boldsymbol{\varphi}(t)}
&& \textstyle \sum \limits_{k = 1}^{\emph{K}} {Y}_{k}(t) \varphi_{k}(t) + \textstyle \sum \limits_{s = 1}^{\emph{S}} {D}_{s}(t) {\varphi}_{s}(t) -  \nu    f_{0}(\boldsymbol{\varphi}(t)), \\
& \text{subject to}
&& \varphi_{k}(t) \leq r_k^{\text{max}}(t).
\end{align}
\end{subequations}
Since the above optimization problem is convex, let $\varphi_{k}^{\ast}(t)$ be the optimal solution obtained by the first order derivative of the objective function of (\ref{eq:Optimal-AR}). With a logarithmic utility function, we have:
\[ \varphi_{k}^{\ast}(t) =  \begin{cases} \frac{\nu \omega_k(t)}{{Y}_{k}(t)} & \quad \text{if } k \leq \emph{M},\\
 \frac{\nu \omega_k(t)}{{Y}_{k}(t) + {D}_{k-\emph{M}}(t)} & \quad \text{otherwise.} \end{cases}\]
The optimal auxiliary variable is $\min \{\varphi_{k}^{\ast}(t), r_k^{\text{max}}(t)\}$.

\subsection{Step 3: Interference Mitigation and Power Allocation}
\label{IM-WB}
For given scheduled users in~\emph{Step~1}, the precoder $\mathbf{U}$ is found by solving~(\ref{zero-ICI}). Finally, problem~(\ref{eq:Obj-Formulate-1}) is decomposed to find the transmit power $p_k^{(b_0)}(t)$ as follows:

\begin{subequations}\label{WB0}
\begin{align}
& \min_{\mathbf{p}(t)}
&& - \textstyle \sum _{k = 1}^{\emph{K}} {A}_{k}(t)  r_{k}(\mathbf{p}(t)), \\
& \text{subject to}
&&  \frac{1}{\emph{N}} \textstyle \sum _{k = 1}^{\emph{K}} \frac{ {p}_{k}^{(b_0)}(t) }{ \Omega_{k}(t) } - P^{(b_0)} \leq 0, \\
&&& {p}_{k}^{(b_0)}(t) \geq 0, \forall k \in \mathcal{K}.
\end{align}
\end{subequations}
The objective function~(\ref{WB0}) is rewritten as $n(\mathbf{p}(t)) = - \textstyle \sum _{k = 1}^{\emph{K}} A_{k}(t) \log \big( 1 +   p_{k}^{(b_0)}(t) n_k(t)  \big)$, where $n_k(t) = \frac{\us_k(t) (1 - \tau_k^2)} {1 +  \textstyle \sum _{s = 1}^{\emph{S}} \op^{(b_s)}(t) \EP_{k}^{(b_s)} (t)}$. The objective function is strictly convex for ${p}_{k}^{(b_0)}(t) \geq 0, \forall k \in \mathcal{K}$, and the constraints are compact. Hence, the optimal solution of $\mathbf{p}^{\star}(t)$ exists, the Lagrangian function is written as $\mathcal{L}(\mathbf{p}(t), \mu_{0}) = n(\mathbf{p}(t)) + \mu_{0} \mathbf{g}(\mathbf{p}(t))$, where $\mu_0 \geq 0$ is the KKT multiplier. The KKT conditions are
\begin{equation}\label{KKT1}
\nabla  n(\mathbf{p}(t))^{T} + \mu_{0} \textstyle  \frac{1}{\emph{N}} \textstyle \sum _{k = 1}^{\emph{K}} \frac{1} { \Omega_{k}(t) }  = 0.
\end{equation}
\begin{equation}
\mu_{0} \Big(\textstyle  \frac{1}{\emph{N}} \textstyle \sum _{k = 1}^{\emph{K}} \frac{ {p}_{k}^{(b_0)}(t)} { \Omega_{k}(t) } - P^{(b_0)} \Big) = 0. \label{KKT2}
\end{equation}
\begin{equation}
 \frac{1}{\emph{N}} \textstyle \sum _{k = 1}^{\emph{K}} \frac{ {p}_{k}^{(b_0)}(t) }{ \Omega_{k}(t) } - P^{(b_0)} \leq 0. \label{KKt3}
 \end{equation}
\begin{equation}
- \mathbf{p}(t) \leq 0, \mu_{0} \geq 0. \label{KKt4}
\end{equation}
Here, $\nabla  n(\mathbf{p}(t))^{T} = (n'({p}_{1}^{(b_0)}(t)), \ldots, n'({p}_{\emph{K}}^{(b_0)}(t)))$ where $n'({p}_{k}^{(b_0)}(t))=  \frac{-A_{k}(t)  n_k(t)}{1 + p_k^{(b_0)}(t)  n_k(t)}$. Since $\mu_{0} \neq 0$, from (\ref{KKT1}), we have

\begin{equation}\label{KKTS1}
p_k^{(b_0)}(t) = \mathrm{max}[\frac{A_k \emph{N} \Omega_{k}(t) }{ \mu_{0}} - \frac{1}{n_k(t)}, 0],
\end{equation}
then, from (\ref{KKT2}) and (\ref{KKTS1}) we derive $\mu_{0}$. Finally, the optimal value of $p_k(t)^{(b_0)\star}$ is obtained with (\ref{KKTS1}).

\subsection{Queue Update}
\label{QU}
Update the virtual queues ${Y}_k(t)$ and ${D}_s(t)$ according to~(\ref{queueY}) and (\ref{queueD}), and the actual queue ${Q}_k(t)$ in~(\ref{queueQ}).
\section{Numerical Results}
\label{Evaluation}
We consider a HetNet scenario, where a MBS is located at the center of the cell. The path loss is modeled as a distance-based path loss with line-of-sight (LOS) model for urban environments at $28$ GHz, $10$ GHz, and $2.4$ GHz~\cite{mW2014}. We denote our proposed algorithms for HetNet (resp. Homogeneous network) as HetNet-Hybrid (resp. HomNet). Here, HomNet refers to when the MBS serves both MUEs and SUEs without SCs. The data arrivals follow the Poisson distribution with the mean rate of $1$ Gbps, $100$ Mbps, and $20$ Mbps for $28$ GHz, $10$ GHz, and $2.4$ GHz, respectively. We consider the proportional fairness utility function, i.e., $f( \bar{r}_{k} ) = \log{\bar{r}_{k}}$~\cite{mo2000fair}. The parameter settings are summarized in~Table~\ref{parameter}.

\begin{table}[t]
\caption{Parameter Settings} 
\centering
\begin{tabular}{ |l|l|l| }
\hline 
Path Loss Model~\cite{mW2014} & Values in dB & Bandwidth (BW) in MHz\\
\hline 
LOS @ 2.4 GHz        & $17 + 37.6 \log(d)$     & 20\\
LOS @ 10 GHz         & $55.25 + 18.5 \log(d)$     & 100\\
LOS @ 28 GHz         & $61.4 + 20 \log(d)$     & 1000\\
\hline
\multicolumn{2}{ |c| }{Parameter} & Values\\ 
\hline 
\multicolumn{2}{ |c| }{ Maximum transmit power of MBS $P^{(b_0)}$} & 43~dBm \\
\multicolumn{2}{ |c| }{ Maximum transmit power of SC}              & 23~dBm \\
\multicolumn{2}{ |c| }{ FD interference threshold $\epsilon_{o}$}  & $5 \times 10^{-3}$ \\
\multicolumn{2}{ |c| }{ Channel estimate error  $\tau$}                   & 0.1      \\
\multicolumn{2}{ |c| }{ SC Antenna Gain }                          & 5 dBi \\
\multicolumn{2}{ |c| }{ Lyapunov parameter  $\nu$}                 & 2 $\times$ $10^3$/1 MHz BW  \\
\multicolumn{2}{ |c| }{ RZF parameter  $\alpha$ }                  & $10^{-2}$\\
\hline
\end{tabular}
\label{parameter}
\end{table}

\begin{figure}[!h]
\begin{minipage}[t]{1\linewidth}
    \centering
   \includegraphics[scale=0.38]{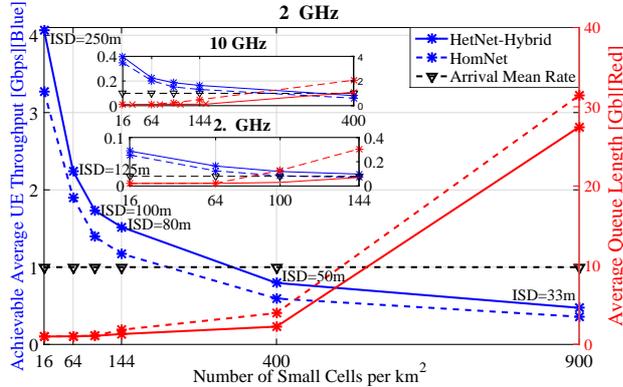} 
   \caption{Achievable Average UE throughput and Network Queue length versus number of Small Cells at $28$ GHz, $10$ GHz, and $2.4$ GHz.}\label{CUE}
\end{minipage}
\end{figure}

\begin{figure}[!h]
\begin{minipage}[t]{1\linewidth}
    \centering
    \includegraphics[scale=0.38]{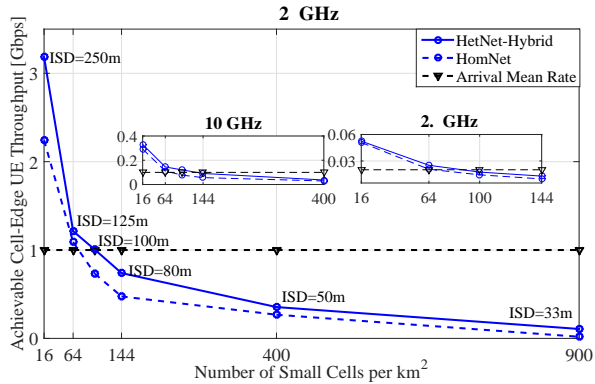} 
    \caption{Achievable $5^{\text{th}}\%$ UE throughput versus number of Small Cells at $28$ GHz, $10$ GHz, and $2.4$ GHz.}
    \label{5UE}
\end{minipage}
\end{figure}

Fig.~\ref{CUE} and~\ref{5UE} report the achievable average UE throughput, cell-edge UE throughput, and average network queue length as a function of network density at different frequency bands. The number of SCs $\emph{S}$ increases from $16$ to $900$, reflecting the inter-site distance (ISD) between SCs decreasing from $250$ m to $33$ m. On the other hand, the number of UEs $\emph{K}$ increases from $32$ UEs to $1800$ UEs per $\text{km}^{2}$. Moreover, the number of antennas $\emph{N}$ at the MBS is twice the number of UEs $\emph{K}$.

With the increase in the number of SCs $\emph{S}$ and MUEs $\emph{M}$, the per UE/SC transmit power is reduced by a factor of ${\emph{K}}$ and the cross-tier interference becomes dominant. Hence, the achievable average UE throughput gradually decreases, while the network queue traffic becomes more congested. Moreover, the UE throughput at $28$ GHz achieves $\textbf{56}\times$ and $\textbf{62}\times$ gain as compared to $2.4$ GHz due to $50\times$ larger bandwidth and smart beamforming, when the ISD is $250$ m and $80$ m, respectively. In ultra-dense deployment, the UE throughput at $2.4$ GHz is below $10$ Mbps, whereas $473$ Mbps per UE is achieved by using $28$ GHz for $33$ m of ISD.

By taking advantage of mmWave frequency bands, we observe that at $28$ GHz when the number of UEs per $\text{km}^{2}$ is increased to $200$ UEs, the average UE throughput reaches $\textbf{1.73 Gbps}$ and $\textbf{1.39 Gbps}$ in case of $\textbf{HetNet-Hybrid}$ and $\textbf{HomNet}$, respectively. Whereas the cell-edge UE throughput reaches $\textbf{1 Gbps}$ and $\textbf{0.73 Gbps}$ in case of $\textbf{HetNet-Hybrid}$ and $\textbf{HomNet}$, respectively. When the ISD is less than $50$ m, the network queue size dramatically increases, and the network becomes congested. To handle this problem, the arrival traffic needs to be adjusted by reducing the admitting data or the UE throughput should increase by increasing the maximum BS transmit power and number of antennas.

\begin{figure}[!h]
\begin{minipage}[t]{1\linewidth}
    \centering
    \includegraphics[scale=0.39]{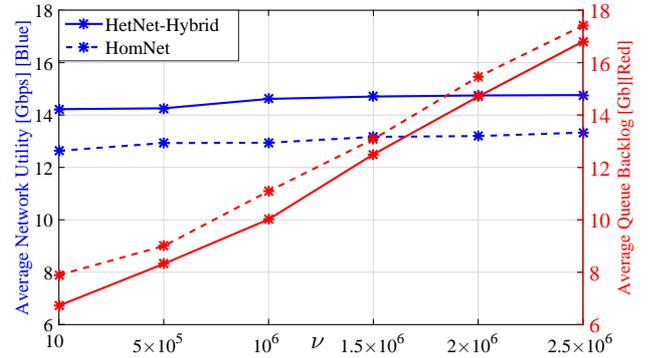}
    \caption{Impact of $\nu$ on the Utility and Network Backlogs at $28$ GHz when $\emph{K} = 16$, $\emph{N} = 64$.}
    \label{Impact_V}
\end{minipage}
\end{figure}

The performance of NUM based on Lyapunov framework is analyzed in~\cite{neely2010S}. There exists an $[\mathcal{O}(1/\nu), \mathcal{O}(\nu)]$ utility-queue backlog tradeoff, which leads to an utility-delay tradeoff. We show the impact of the Lyapunov parameter $\nu$ on the achievable average network utility and queue backlog as seen in Fig.~\ref{Impact_V}, when $\emph{K} = 16$, $\emph{N} = 64$, and $P^{(b_0)} = 38$ dBm. By varying the value of $\nu$, the network utility is increasing with $\mathcal{O}(1/\nu)$, while the network backlog linearly increases with $\mathcal{O}(\nu)$.
\section{~Conclusion}
\label{Conclusion}
In this paper, we have studied the problem of joint in-band backhaul scheduling and interference mitigation in $5$G HetNets which takes into account the combination of in-band FD-SC and massive MIMO technologies. The main addressed problem is to simultaneously schedule macro cell users and to provide the wireless backhaul to small cells. The goal is to maximize a network utility function of the total time-average data rates subject to the wireless backhaul constraint and network stability in the presence of imperfect CSI. At mmWave $28$ GHz with $1$ GHz of BW, we observe that $\textbf{1.7 Gbps}$ of the achievable average UE throughput and $\textbf{1 Gbps}$ of the cell-edge UE throughput can be reached for $200$ UE per $\text{km}^{2}$ in HetNet. When the network gets dense, harnessing mmWave yields  $\textbf{62}\times$ gain of UE throughput as compared to conventional cellular frequency.

\appendix[Proof of Closed-Form Expression]
\label{Appendix1}
By applying the technique in~\cite[Theorem 2, 3]{wagner2012l}, the deterministic equivalent of the asymtotic $\mathrm{SINR}$ of the MUE $m$ is
\begin{align} \label{eq:dl-sinr-MUE-rzf-0}
\gamma_{m}^{(b_0)} &\xrightarrow{a. s.} \frac { \us_{m} p_{m}^{(b_0)} (1 - \tau_m^2) (\Omega_{m}^{})^2}{ \Phi }, \nonumber
\end{align}
where $\Phi = \Upsilon_{m}^{}\Big[ \alpha^2 -  \tau_m^2 \big(\alpha^2 - (\alpha + \Omega_{m})^2 \big)\Big] +
(\alpha + \Omega_{m}^{})^2(1 + \textstyle \sum _{s = 1}^{\emph{S}} \op^{(b_s)} \EP_{m}^{(b_s)})$. $\Omega_{m} =  \frac{1}{N} \text{Tr} (\mathbf{\tilde{\Theta}}_m \mathbf{G})$ forms the unique positive solution of which is the Stieltjes transform of nonnegative finite measure, where  $\textstyle \mathbf{G} =  \Big( \frac{1}{N} \sum _{k = 1}^{\emph{K}} \frac{ \mathbf{\tilde{\Theta}}_k }{ \alpha + \Omega_{k}^{}} + \mathbf{I}_{N_i} \Big)^{-1}$. Here, $\textstyle \Upsilon_{m}^{} =  \frac{1}{N} \sum _{k = 1, k \neq m}^{\emph{K}} \frac{ \alpha^2 \us_{k} p_{k}^{(b_0)} e_{km}}{
(\alpha + \Omega_{k}^{})^2 }$, and,
where $\mathbf{\tilde{\Theta}}_k = \mathbf{U} \mathbf{U}^{\dag} \mathbf{\Theta}_k^{(b_0)} \mathbf{U} \mathbf{U}^{\dag}$. $\mathbf{e} = [e_{k}], k \in \mathcal{K}$, and $\mathbf{e_m} = [e_{mk}], k \in \mathcal{K}$ are given by $\mathbf{e} = ( \mathbf{I} - \mathbf{J})^{-1}\mathbf{u}$, $\mathbf{e_k} = ( \mathbf{I} - \mathbf{J})^{-1}\mathbf{u_k}$, where $\mathbf{J} = [J_{ij}], i,j \in \mathcal{K}$. $\mathbf{u} = [u_{k}], k \in \mathcal{K}$, $\mathbf{u_m} = [u_{mk}], k \in \mathcal{K}$ are given by $\mathbf{J}_{ij} = \displaystyle  \frac{ \frac{1}{N} \text{tr} { \mathbf{\tilde{\Theta}}_i \mathbf{G} \mathbf{\tilde{\Theta}}_j \mathbf{G}} } {N(\alpha + \Omega_{j}^{})^2}$, $u_{mk} = \frac{1}{\alpha^2 N} \text{tr} { \mathbf{\tilde{\Theta}}_k \mathbf{G}  \mathbf{\tilde{\Theta}}_m \mathbf{G}}$, $u_{k} = \frac{1}{\alpha^2 N} \text{tr} { \mathbf{\tilde{\Theta}}_k \mathbf{G}^2}$. Similarly, the SINR of the SC $b_s$ are
\begin{align}
\gamma_{s}^{(b_0)} &\xrightarrow{a.s.}  \frac { \us_{s} p_{s}^{(b_0)} (\Omega_{s}^{})^2}  { \alpha^2
\Upsilon_{s}^{} + (\alpha + \Omega_{s}^{})^2(1 + \sum _{s' = 1, s' \neq s}^{\emph{S}} \op^{(b_{s'})} \EP_{s}^{(b_{s'})}) }. \nonumber
\end{align}

\bibliographystyle{IEEEtran}
\bibliography{SmallCellRef}

\begin{thebibliography}{10}
\providecommand{\url}[1]{#1}
\csname url@samestyle\endcsname
\providecommand{\newblock}{\relax}
\providecommand{\bibinfo}[2]{#2}
\providecommand{\BIBentrySTDinterwordspacing}{\spaceskip=0pt\relax}
\providecommand{\BIBentryALTinterwordstretchfactor}{4}
\providecommand{\BIBentryALTinterwordspacing}{\spaceskip=\fontdimen2\font plus
\BIBentryALTinterwordstretchfactor\fontdimen3\font minus
  \fontdimen4\font\relax}
\providecommand{\BIBforeignlanguage}[2]{{%
\expandafter\ifx\csname l@#1\endcsname\relax
\typeout{** WARNING: IEEEtran.bst: No hyphenation pattern has been}%
\typeout{** loaded for the language `#1'. Using the pattern for}%
\typeout{** the default language instead.}%
\else
\language=\csname l@#1\endcsname
\fi
#2}}
\providecommand{\BIBdecl}{\relax}
\BIBdecl

\bibitem{Nokia2011}
``2020: Beyond 4g radio evolution for the gigabit experience,'' White Paper,
  Noikia Siements Networks, 2011.

\bibitem{marzt2010non}
T.~L. Marzetta, ``Noncooperative cellular wireless with unlimited numbers of
  base station antennas,'' \emph{IEEE Transactions on Wireless Communications},
  vol.~9, no.~11, pp. 3590--3600, 2010.

\bibitem{li2015small}
B.~Li, D.~Zhu, and P.~Liang, ``Small cell in-band wireless backhaul in massive
  multiple-input multiple-output systems,'' in \emph{IEEE International
  Conference on Communications}, 2015.

\bibitem{S2014if}
L.~Sanguinetti, A.~Moustakas, and M.~Debbah, ``{Interference management in 5G
  reverse TDD HetNets: A large system analysis},'' \emph{IEEE Journal on
  Selected Areas in Communications}, vol.~33, pp. 1187--1200, 2015.

\bibitem{rusek2013s}
F.~Rusek, D.~Persson, B.~Lau, E.~Larsson, T.~Marzetta, O.~Edfors, and
  F.~Tufvesson, ``{Scaling up MIMO: Opportunities and challenges with very
  large arrays},'' \emph{IEEE Signal Processing Magazine}, vol.~30, no.~1, pp.
  40--60, 2013.

\bibitem{wagner2012l}
S.~Wagner, R.~Couillet, M.~Debbah, and D.~Slock, ``{Large system analysis of
  linear precoding in correlated MISO broadcast channels under limited
  feedback},'' \emph{IEEE Transactions on Information Theory}, vol.~58, no.~7,
  pp. 4509--4537, 2012.

\bibitem{Liu2014}
A.~Liu and V.~Lau, ``{Hierarchical Interference Mitigation for Massive MIMO
  Cellular Networks},'' \emph{IEEE Transactions on Signal Processing}, vol.~62,
  no.~18, pp. 4786--4797, Sept 2014.

\bibitem{pareto2011}
H.~Boche, S.~Naik, and M.~Schubert, ``Pareto boundary of utility sets for
  multiuser wireless systems,'' \emph{IEEE/ACM Transactions on Networking},
  vol.~19, no.~2, pp. 589--601, 2011.

\bibitem{pareto2012}
Z.~Chen, S.~Vorobyov, C.~Wang, J.~Thompson \emph{et~al.}, ``{Pareto region
  characterization for rate control in MIMO interference systems and Nash
  bargaining},'' \emph{IEEE Transactions on Automatic Control}, vol.~57,
  no.~12, pp. 3203--3208, 2012.

\bibitem{neely2010S}
M.~J. Neely, ``Stochastic network optimization with application to
  communication and queueing systems,'' \emph{Synthesis Lectures on
  Communication Networks}, vol.~3, no.~1, pp. 1--211, 2010.

\bibitem{beck2010seq}
A.~Beck, A.~Ben-Tal, and L.~Tetruashvili, ``A sequential parametric convex
  approximation method with applications to nonconvex truss topology design
  problems,'' \emph{Journal of Global Optimization}, vol.~47, no.~1, pp.
  29--51, 2010.

\bibitem{li2014e}
H.~Li, L.~Song, and M.~Debbah, ``Energy efficiency of large-scale multiple
  antenna systems with transmit antenna selection,'' \emph{IEEE Transactions on
  Communications}, vol.~62, no.~2, pp. 638--647, 2014.

\bibitem{mW2014}
M.~R. Akdeniz, Y.~Liu, M.~K. Samimi, S.~Sun, S.~Rangan, T.~S. Rappaport, and
  E.~Erkip, ``Millimeter wave channel modeling and cellular capacity
  evaluation,'' \emph{IEEE Journal on Selected Areas in Communications},
  vol.~32, no.~6, pp. 1164--1179, 2014.

\bibitem{mo2000fair}
J.~Mo and J.~Walrand, ``Fair end-to-end window-based congestion control,''
  \emph{IEEE/ACM Transactions on Networking (ToN)}, vol.~8, no.~5, pp.
  556--567, 2000.

\end{thebibliography}
\end{document}